\documentclass[twocolumns,structabstract]{aa}
\usepackage{amsmath,amsfonts,amstext}
\usepackage{amssymb}
\usepackage{graphicx}
\usepackage{txfonts}
\usepackage{natbib}
\usepackage[english]{babel}

\bibpunct{(}{)}{;}{a}{}{,} 
\def\icarus{{Icarus }}           

\def\apjs{{Astrophys. J. Suppl. }}

\def\apjl{{Astrophys. J. (Letters) }}
\def\planss{{Planet. Space Sci. }}
\def\apss{{Astrophys. and Space Sci. }}
\def\aap{{Astron. and Astrophys. }}

\def\pra{{Phys. Rev. A }}
\def\jgr{{J. Geophys. Res. }}

\def\ssrs{{Space Science Reviews }}

\def\jcp{{J. Chem. Phys. }}

\begin{document}

\title{Heating efficiency in hydrogen-dominated upper atmospheres}
\titlerunning{Heating efficiency in hydrogen-dominated upper atmospheres}

   \author{V. I. Shematovich,
          \inst{1}
          D.E.~ Ionov,
          \inst{1}
          \and
          H.~Lammer
          \inst{2}}
\authorrunning{Shematovich et al.}
\offprints{V. I. Shematovich,\\
\email{shematov@inasan.ru}}

   \institute{Institute of Astronomy, Russian Academy of Sciences, 48 Pyatnitskaya str., Moscow, 119017 Russian Federation
   \and
    Space Research Institute, Austrian Academy of Sciences, Schmiedlstrasse 6, A-8042 Graz, Austria}

\date{Received \today}

\abstract
{The heating efficiency $\eta_{\rm h\nu}$ is defined as the ratio of the net local gas-heating rate
to the rate of stellar radiative energy absorption. It plays an important role in thermal-escape processes 
from the upper atmospheres of planets that are exposed to stellar soft X-rays and extreme ultraviolet radiation (XUV).}
{We model the thermal-escape-related heating efficiency $\eta_{\rm h\nu}$ of the stellar XUV radiation in the hydrogen-dominated upper atmosphere of the extrasolar gas giant HD 209458b. The model result is then compared with previous thermal-hydrogen-escape studies which assumed $\eta_{\rm h\nu}$ values between 10--100\%.}
{The photolytic and electron impact processes in the thermosphere were studied by solving the kinetic Boltzmann equation and applying a Direct Simulation Monte Carlo model. We calculated the energy deposition rates of the stellar XUV flux and that of the accompanying primary photoelectrons that are caused by electron impact processes in the H$_2$ $\rightarrow$ H transition region in the upper atmosphere.}
{The heating by XUV radiation of hydrogen-dominated upper atmospheres does not reach higher than 20\% above the main thermosphere altitude, if the participation of photoelectron impact processes is included.}
{Hydrogen-escape studies from exoplanets that assume $\eta_{\rm h\nu}$ values that are $\geq$ 20 \% probably overestimate the thermal escape or mass-loss rates, while those who assumed values that are $<$ 20\% probably produce more realistic atmospheric-escape rates.}

\keywords{exoplanets -- XUV -- thermosphere -- heating efficiency  -- atmospheric escape}

\maketitle

\section{Introduction}
More than 1000 exoplanets are known today (http://exoplanets.eu), and the detection of hydrogen- and volatile-rich exoplanets 
at orbital distances $<$1 AU raises questions about their upper atmospheric structures and the stability against escape of atmospheric gases. 
Since $\geq 40$ per cent of all discovered exoplanets are orbiting their host stars at distances closer than the orbit of Mercury, the atmospheres of 
these bodies evolve in much more extreme environments than what is currently known from the planets in our solar system. Therefore, more intense stellar X-ray, soft X-ray, and extreme ultraviolet radiation (XUV: $\lambda \sim$1--100 nm) and particle fluxes at these close orbital distances will strongly change the upper atmospheric structure of these objects. To some extent the radiation fluxes that expose the upper atmospheres of close-in exoplanets can be considered to be similar to the XUV flux levels of the young Sun after its arrival at the zero-age main sequence \citep{Ribas-et-al-2005, Claire-et-al-2012}.

The photolysis of hydrogen-dominated upper atmospheres of the close-in exoplanets by the XUV radiation of the parent star lead to the formation of suprathermal particles (i.e., particles with an excess of kinetic energy), primary photoelectrons from ionization of H$_2$, He, and H atoms resulting from dissociation, and dissociative ionization processes  of H$_2$. These particles with excess kinetic energies are an important source of thermal energy in the upper atmosphere of hydrogen-rich planets.

\cite{Penz-et-al-2008} studied the XUV-driven hydrodynamic hydrogen escape from the hot Jupiter HD 209458b
over its evolutionary time period and found that the thermal mass-loss rate can be approximated by a modified energy-limited formula that includes
a mass-loss enhancement factor due to a Roche-lobe effect \citep{Erkaev-et-al-2007} and a heating efficiency $\eta_{\rm h\nu}$ for the stellar XUV radiation. The heating efficiency $\eta_{\rm h\nu}$ can be defined as the ratio of the net local gas-heating rate
to the rate of stellar radiative energy absorption. From early studies of thermal escape from hydrogen-dominated primordial atmospheres of accreting protoplanets \citep{Zahnle-et-al-1988} to recent XUV-powered hydrogen-mass-loss studies from exoplanets \citep{Lanza-2013, Wu-and-Lithwick-2013}  the heating efficiency $\eta_{\rm h\nu}$ of the hydrogen gas has been neglected or assumed to be within $\sim$10--100\%. Morevover, many hydrodynamic atmospheric escape studies from hot Jupiters and other expected hydrogen-dominated exoplanets are based on a total conversion of the absorbed stellar XUV energy into energy that powers the escape.

A heating efficiency value $\eta_{\rm h\nu}$ of 100 \% has been assumed in the thermal escape studies by \cite{Lammer-et-al-2003},  \cite{Baraffe-et-al-2004},  \cite{Lecavelier-des-Etangs-2004},  \cite{Hubbard-et-al-2007a},  \cite{Hubbard-et-al-2007b},  \cite{Lecavelier-des-Etangs-2007},  \cite{Davis-and-Wheatley-2009},  \cite{Sanz-Forcada-et-al-2010},  \cite{Lissauer-et-al-2011},  \cite{Sanz-Forcada-et-al-2011},  \cite{Lanza-2013}, and \cite{Wu-and-Lithwick-2013}. Recently, \cite{Kawahara-et-al-2013} assumed a heating efficiency $\eta_{\rm h\nu}$ of 50\% to study the hydrogen evaporation of the planet candidate KIC 12557548b. 

More detailed studies such as that by \cite{Yelle-2004}, who applied a 1D hydrodynamical upper atmosphere model 
that considers hydrogen photochemistry in the thermosphere of hot Jupiters between 0.01--0.1 AU found that $\eta_{\rm h\nu}$ 
is $\sim$40--60\% at planetary distances of $\sim$1.03--1.05$R_{\rm p}$, $\sim$20\% around $\sim$1.4$R_{\rm p}$, and
$\sim$15\% at distances $>1.4R_{\rm p}$. 
From this result several researchers assumed an average $\eta_{\rm h\nu}$ value of 30\% in their studies, which also agrees with \cite{Watson-et-al-1981}, who applied a similar height-integrated average $\eta_{\rm h\nu}$ value of 30\% for the thermal-escape studies of a hydrogen-rich early Earth. These $\eta_{\rm h\nu}$ values are close to the $\sim$15--30\% estimated by \cite{Chassefiere-1996} for the study of hydrodynamic escape of hydrogen from a H$_2$O-rich upper atmosphere of early Venus.

Other studies assumed different $\eta_{\rm h\nu}$ values in their XUV-powered thermal-escape studies, which clearly shows that the heating efficiency may affect the thermal escape for a particular planet or during its XUV-flux driven evolution. \cite{Murray-Clay-et-al-2009} used an efficiency of ~30\% for HD 209458b, but they found lower efficiencies at higher incident fluxes (e.g., ~10\% for TauTauri-like fluxes). The variability is expained by radiative losses. \cite{Penz-et-al-2008} assumed for thermal-escape studies of the hot Jupiter HD 209458b $\eta_{\rm h\nu}$ values of 10\%, 60\%, and 100\%, \cite{Lammer-et-al-2009} applied $\eta_{\rm h\nu}$ values for mass-loss studies of exoplanets with known size and mass of 10\%, 25\%, 60\%, and 100\%, \cite{Jackson-et-al-2010} studied the mass loss from CoRoT-7b and assumed $\eta_{\rm h\nu}$ values of 10\%, 25\%, 50\%, and 100\%. One year later, \cite{Leitzinger-et-al-2011} assumed
$\eta_{\rm h\nu}$=25\% for mass-loss studies from CoRoT-7b and Kepler-10b, \cite{Ehrenreich-and-Desert-2011} studied the thermal mass-loss evolution of close-in exoplanets by assuming $\eta_{\rm h\nu}$ values of 1\%, 15\%, and 100\%, and \cite{Jackson-et-al-2012} investigated the X-ray heating contribution and assumed for the XUV heating efficiency $\eta_{\rm h\nu}$ a lower value of 25\% and the energy-limited approach of 100\%. \cite{Koskinen-et-al-2013} studied the escape of heavy atoms from the ionosphere of HD 209458b with a photochemical-dynamical thermosphere model for various $\eta_{\rm h\nu}$ values of 10\%, 30\%, 50\%, 80\%, and 100\%.

Since the past two years, several studies such as that of \cite{Lopez-et-al-2012} or \cite{Lopez-and-Fortney-2013} assumed lower $\eta_{\rm h\nu}$ values of between 10--20\% for the thermal evolution and mass loss of  super-Earth and sub-Neptune planets in the Kepler-11 system, and studied the role of the core mass in the evaporation of the Kepler radius distribution and the Kepler-36
density dichotomy. \cite{Kurokawa-and-Kaltenegger-2013} applied a heating efficiency $\eta_{\rm h\nu}$ of 25\% to their mass-loss study of CoRoT-7b and Kepler-10b  similar to \cite{Leitzinger-et-al-2011}. \cite{Valencia-et-al-2013} studied the bulk composition and thermal escape of the super-Earth GJ 1214b and other sub-Neptune-type exoplanets by assuming a lowest $\eta_{\rm h\nu}$ value of 10\% and a highest value of 40\%. More or less similar lowest and highest $\eta_{\rm h\nu}$ values of 15\% and 40\% have been assumed in recent works by \cite{Erkaev-et-al-2013}, \cite{Lammer-et-al-2013}, \cite{Erkaev-et-al-2014}, \cite{Kislyakova-et-al-2013, Kislyakova-et-al-2014}, and \cite{Lammer-et-al-2014}, who studied the escape of hydrogen envelopes from early Mars and sub-Earth to super-Earths inside the habitable zone of a solar-like G-type star for XUV fluxes that are higher than several times to up to 100 times of the present-day Sun, as well as for five exoplanets between the super-Earth and mini-Neptune domain in the Kepler-11 system. Finally, in their recent study on impact-related photoevaporative mass-loss on masses and radii of H$_2$O-rich sub- and super-Earths, \cite{Kurosaki-et-al-2013} assumed an $\eta_{\rm h\nu}$ value of 10\%. The lower value of these studies of 15\% was also chosen by \cite{Kasting-and-Pollack-1983} in their pioneering study on the hydrodynamic escape of a water-rich early atmosphere of Venus.

From this brief overview of assumed heating efficiency values $\eta_{\rm h\nu}$ between 10--100\% it is clear that by assuming an incorrect value, one can over- or underestimate the thermal escape rates within an order of magnitude. Therefore, it is timely to assign the realistic fraction of stellar XUV radiation that is transformed into the heating of upper atmospheres of hydrogen-dominated planets to evaluate how deep in an atmosphere, for instance,  in the XUV spectral range, stellar photons continue to release part of their energy as heat into the surrounding neutral gas. Thus, the aim of this work is a detailed study of the XUV-related heating efficiency in hydrogen-rich upper atmospheres. In Sect. 1 we discuss photolytic and electron impact processes in a hydrogen-dominated thermosphere. In Sect. 2 we describe a Direct Simulation Monte Carlo (DMSC) model that we used to study
the photoelectron movement in the background atmosphere, collisions, and energy distribution. Finally, we present the results in Sect. 3 and discuss our findings in comparison with the previous studies.

\section{Heating efficiency modeled in hydrogen-dominated upper atmospheres}
To estimate the effect of the XUV emission from solar-type stars of different ages an accurate description of radiative transfer
and photoelectron energy deposition is required. ~\cite{Cecchi-Pestellini-2006, Cecchi-Pestellini-2009} showed that X-rays strongly contribute to the heating of hydrogen-dominated planetary atmospheres of close-in exoplanets. The flux of stellar XUV emission photons incident upon a planetary atmosphere of hydrogen-dominated composition photoionizes the gas so that a flux of high-energy photoelectrons can be produced, which again deposit their energy into the gas.

In a partially neutral gas, electrons ionize, excite, and dissociate atomic and molecular species, as well as heat the gas through Coulomb collisions.
In determining these energy depositions, we must account for all the possible degradation histories of the energetic electrons. When the stopping
medium is only partially neutral, electron-electron interactions contribute to the electron energy degradation, and a significant portion of the energetic
electron energy is deposited into the stopping medium as heat. As the ionized fraction rises, more and more of the electron energy heats the gas,
while the excitation and ionization yields decrease. In the following sections we model the heating efficiency $\eta_{\rm h\nu}$ in the hydrogen-dominated upper atmosphere of the well-studied hot Jupiter HD 209458b. The results can also be used for any thermal-escape study of hydrogen-rich upper atmospheres.

\subsubsection{Photolytic and electron-impact processes in the upper atmosphere}
The incoming stellar XUV flux decreases because of absorption in the upper atmosphere, which results in dissociation and ionization and, hence,
in heating of the upper atmosphere. The extreme UV radiation of the star is absorbed by the atmospheric gas and leads to excitation,
dissociation, and ionization of different components of the atmosphere. For an atmosphere dominated by H$_2$, H, and He, the following photolytic
processes have to be taken into account:
\begin{equation}
{\rm H_2 +h\nu,(e_{\rm p}) \rightarrow \left\{
\begin{array}{l}
 {\rm H(1s)+H(1s,2s,2p)+(e_{\rm p})}\\
{\rm H_2^+ + e+ (e_{\rm p})} \\
{\rm  H(1s)+H^++e+(e_{\rm p})}\end{array}\right.}
\label{1eq}
\end{equation}
\begin{equation}
{\rm H,He + h \nu, (e_{\rm p}) \rightarrow H^+,He^+ + e + (e_{\rm p}),}
\label{2eq}
\end{equation}
where e$_{\rm p}$ is a photoelectron.  The photoionization processes~\eqref{1eq} and~\eqref{2eq} produce photoelectrons with energies sufficient for the subsequent ionization and excitation of atomic and molecular hydrogen. The energy of the ionizing quanta by definition exceeds the ionization potential, and its excess produces electrons with an excess of kinetic energy and ions in excited states. The differential photoelectron
production rate $q_{\rm e}(E,r)$ per volume at a given altitude $r$ in the upper atmosphere can be defined by the following expression:
\begin{equation}
\begin{array}{l}
 q_{\rm e}(E,r)=\sum_{\rm k} q^{(k)}_{\rm e} (E,r) \\
 q^{(\rm k)}_{\rm e}(E,r) = \sum_l n_{\rm k}(r) \int\limits_0^{\lambda_{\rm k}}d\lambda I_{\rm \infty} (\lambda)exp(-\tau(\lambda,r))\sigma^{\rm i}_{\rm k} p_{\rm k}(\lambda, E_{\rm k,l})
\end{array}
\label{3eq}
\end{equation}
where the optical thickness $\tau$ is given by
\begin{equation}
\tau(\lambda,r)=\sum_{\rm k} \sigma^{\rm a}_{\rm k}(\lambda) \int\limits_r^\infty n_{\rm k}(r')dr',
\label{4eq}
\end{equation}
and $n_{\rm k}$ is the neutral number density of component $k$. $\sigma^i_{\rm k}(\lambda)$ and $\sigma^a_{\rm k}(\lambda)$ are the corresponding
ionization and absorption cross-sections, dependent on the wavelength $\lambda$.
In expression~\eqref{3eq}, we use the relative yields $p_{\rm k}(\lambda,E_{\rm k,l})$ to form ions of species $k$ in the internal excitation state $l$ when neutral species are ionized by the photon with the wavelength $\lambda$, and the potential of ionization $E_{\rm k,l}$ for the electronically excited states $l$ of the ion. The energy of the forming photoelectron is $E=E_{\rm \lambda}- E_{\rm k,l}$, where $E_{\rm \lambda}$ is the energy of the photon and $\lambda_{\rm k}$ is the wavelength corresponding to the ionization potential of the $k_{th}$ neutral component. $I_\infty(\rm \lambda)$ is the number flux of the incident stellar radiation at the wavelength $\lambda$. In formula~\eqref{3eq} term $q^{(\rm k)}_{\rm e}$ represents a partial by neutral species differential production rate of photoelectrons in the photoionization processes.
Because spectra of the stellar XUV fluxes for HD 209458 are currently poorly known~\citep{Lammer-2012}, we used the flux of solar radiation in the wavelength range of 1-115 nm for the moderate-activity solar spectrum model from~\cite{Huebner-1992} scaled to the distance of 0.045 AU equal to the semi-major axis of close-in exoplanets such as HD 209458b. This approach is valid because HD 209458 is a solar-like G-type star with a similar age  as the Sun \citep{Vidal-Madjar-et-al-2003}.

\begin{figure}
\includegraphics[width=0.94\columnwidth]{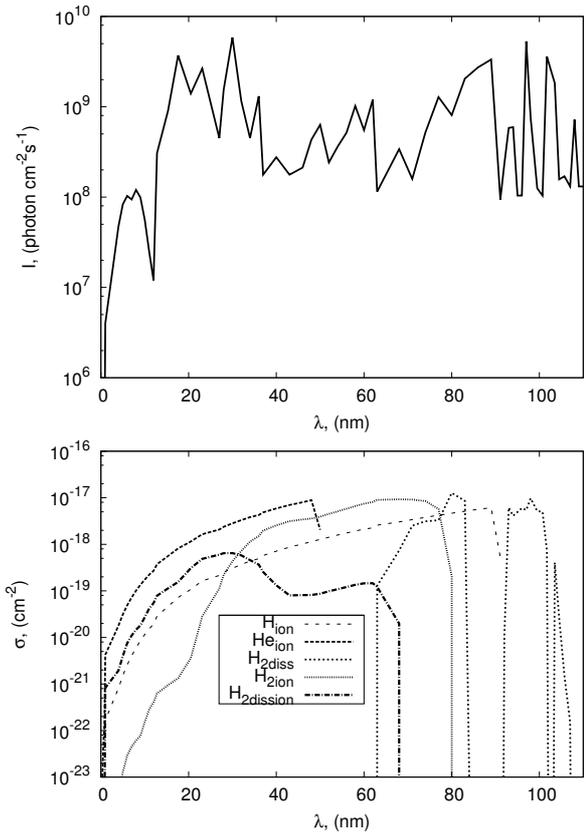}
\caption{
Model input parameters related to (upper panel) the  spectrum of
the solar XUV radiation for a moderate activity level, and (bottom panel) ionization and dissociation cross-sections for the main atmospheric components H$_2$, H, and He \citep{Huebner-1992}.}
\label{fig1}
\end{figure}

The relative yields for excited ionic states, absorption, and ionization cross-sections are also taken from ~\cite{Huebner-1992}
for the main atmospheric components H$_2$, H, and He. The model parameters from \cite{Huebner-1992}
are shown in Fig.~\ref{fig1}.
The newly formed electrons are transported in the thermosphere, where they lose
their kinetic energy in elastic, inelastic, and ionization collisions with the
ambient atmospheric gas
\begin{equation}
e(E)+X \rightarrow \left\{
\begin{array}{l}
 e(E') +X \\
 e(E') + X^* \\
 e(E') + X^+ + e(E_s)
\end{array}
\right\},
\label{5eq}
\end{equation}
where $E$ and $E'(<E)$ are the kinetic energies of the primary electron before
and after a collision, $X=$H$_2$, H, He; $X^*$ and $X^+$ are atmospheric species
in excited and ionized states, $E_{\rm s}$ is the energy of the secondary electron
formed in the ionizing collision. We considered the following neutral excited
states for the main atmospheric species:
\begin{itemize}
\item excitation and dissociative excitation of H$_2^*$=H$_2$ (rot, vib, electronic states A$^3$, B$^3$, C$^3$, B$^1$, C$^1$, E$^1$, B'$^1$, D$^1$, D'$^1$, B''$^1$,
$\Pi_{\rm s}$, Ly-$\alpha$);
\item direct ionization of H$_2\rightarrow$H$_2^+$;
\item dissociative ionization of H$_2\rightarrow$H$^+$ + H;
\item excitation of He$^*=$He (21 electronic states at energies between 20.61
and 23.91 eV);
\item direct ionization of He$\rightarrow $He$^+$;
\item excitation of H$^*=$H (9 states 1s2p - 1s10p);
\item direct ionization of H$\rightarrow$H$^+$.
\end{itemize}
Neutral metal atoms in a hydrogen-rich atmosphere, similarly
as in a dense interstellar cloud, could increase the fractional ionization via charge-transfer reactions of 
molecular ions (Oppenheimer and Dalgarno 1974). If the collision produces ionization, 
a secondary electron is created and is randomly assigned an isotropically distributed pitch angle and an energy, using
an integral form of the formula of~\cite{Green-and-Sawada-1972}
and~\cite{Jackman-1977} based on the laboratory results of~\cite{Opal-1971}
\begin{equation}
\int\limits_0^{E_{\rm s}}
\sigma_{i,j}(E_{\rm p},E')dE'=A(E_{\rm p})\Gamma(E_{\rm p})\left[\tan^{-1}\left(\frac{
E_{\rm s}-T_0(E_{\rm p})}{\Gamma(E_{\rm p})}\right)+c\right],
\end{equation}
where $\sigma_{\rm i,j}(E_{\rm p},E_{\rm s})$  is the state-specific cross-section for species
$i$ and state $j$ at primary electron energy $E_p$ and secondary electron energy
$E_{\rm s}$ , $A(E_{\rm p})$, $\Gamma(E_{\rm p})$, and $T_0(E_{\rm p})$ are fitting functions defined by the tabulated parameters of~\cite{Jackman-1977}, and
$c=\tan^{-1}\left[\frac{T_0(E_{\rm p})}{\Gamma(E_{\rm p})}\right]$. Energy $E_{\rm s}$ of the secondary electron produced by an ionization collision is calculated by solving the equation (6) according to the procedure described by~\cite{Garvey-Green-1976},~\cite{Jackman-1977}, and~\cite{Garvey-1977}.

For inelastic collisions, a forward-scattering
approximation was used: we assumed that the phase function from these
collisions is so strongly peaked in the forward direction that angular
redistribution by this process is negligible. Below energies of 100 eV considerable backscattering
can occur from forbidden excitation transitions,
but the flux becomes so isotropic and the relative size of the
elastic cross-sections becomes so large that this has little effect on the final
pitch-angle distribution.

\subsection{Model description}
\subsubsection{Kinetic equation}
The fresh electrons lose their excess kinetic energy in collisions with the
ambient atmospheric particles. Their kinetics and transport is described by the
kinetic Boltzmann equation~(\cite{Shematovich-2008, Shematovich-2010})
\begin{equation}
  \vec{v}\frac{\partial}{\partial \vec{r}}f_{\rm e} + \frac{\vec{Y}}{m_{\rm e}}
\frac{\partial}{\partial \vec{v}} f_{\rm e} = Q_{e,photo}(\nu)+ Q_{e, secondary}(\nu)+
\sum_{M=H,He,H_2} J(f_{\rm e},f_{\rm M}),
 \label{6eq}
 \end{equation}
where $f_{\rm e}(\vec{r},\vec{v})$, and $f_{\rm M}(\vec{r},\vec{v})$ are the velocity
distribution functions for electrons and for the species of the ambient gas,
respectively. The left side of the kinetic equation describes the transport of
electrons in the planetary gravitational field $\vec{Y}$. In the right-hand side
of the kinetic equation the $Q_{\rm e,photo}$ term describes the formation
rate of primary electrons due to photoionization, while the
$Q_{\rm e,secondary}$ term describes the formation rate of the
secondary electrons. The elastic and inelastic scattering terms $J$ for electron
collisions with ambient atmospheric species are written in a standard form. We
assumed that the ambient atmospheric gas is characterized by the local
Maxwellian velocity distribution functions.

\subsubsection{Numerical model}
The DSMC method is an efficient tool to solve atmospheric kinetic systems in the stochastic
approximation~\citep{Shematovich-1994, Bisikalo-1995, Marov-1996, Gerard-2000}. The details
of the algorithmic realization of the numerical model were given
earlier~\citep{Shematovich-1994, Bisikalo-1995, Shematovich-2010}. In the numerical simulations,
the evolution of the system of modelling particles that is caused by collisional processes
and particle transport is calculated from the initial to the steady state. To minimize boundary effects, the lower boundary is set at altitudes 
where the atmosphere is collision-dominated and the upper boundary is fixed at altitudes
where the atmospheric gas flow is practically collisionless. The relative
importance of the collisional processes is governed by their cross-sections. In
this particular realization of the model, we used experimental and calculated
data for the cross-sections and distributions of the scattering angles in the
elastic, inelastic, and ionization collisions of  electrons with H$_2$, He, and
H taken from the following sources: (a) for electron collisions with H$_2$ we
used the AMDIS database (https://dbshino.nfs.ac.jp) and the work by~\cite{Shyn-Sharp-1981};
and (b) for electron collisions with He and atomic hydrogen, we use the H$_2$ NIST database
(http://physics.nist.gov/PhysRef. Data/Ionization/) and the
data from the studies of~\cite{Jackman-1977} and~\cite{Dalgarno-1999}.

\subsubsection{Energy deposition of the stellar soft X-ray and EUV radiation}
The partial deposition rates of the stellar XUV radiation due to the photolytic
processes~\eqref{1eq} and~\eqref{2eq} in the H$_2 \rightarrow $H transition
region in the upper atmosphere of HD 209458b can be calculated in accordance with
formula~\eqref{3eq} as follows:
\begin{equation}
\begin{array}{l}
 W_{\rm h\nu}(r)=\sum_{\rm k} W^{(k)}_{\rm h\nu} (r) \\
 W^{(k)}_{\rm h\nu} = \sum_{\rm l} n_{\rm k}(r) \int\limits_0^{\lambda_{\rm i}}d\lambda
E_{\rm \lambda}I_\infty (\lambda)exp(-\tau(\lambda,r))\sigma^a_{\rm k}
p_{\rm k}(\lambda, E_{\rm k,l}),
\end{array}
\end{equation}
where $W_{\rm h\nu}(z)$ and $W^{\rm (k)}_{\rm h\nu}(r)$ are the local total and partial
deposition rates of stellar XUV radiation in the upper atmosphere. The rate
$W_{\rm pe}(r)$ of kinetic energy storage in the primary or fresh photoelectrons
is equal to
\begin{equation}
\begin{array}{l}
 W_{\rm pe}(r)=\sum_{\rm k} W^{\rm (k)}_{\rm pe} (r) \\
 W^{\rm (k)}_{\rm pe} = \sum_{\rm l} n_{\rm k}(r) \int\limits_0^{\rm \lambda_i}d\lambda
(E_{\rm \lambda}-E_{\rm k,l})I_\infty (\lambda)exp(-\tau(\lambda,r))\sigma^i_{\rm k}
p_{\rm k}(\lambda, E_{\rm k,l}).
\end{array}
\end{equation}
Using the DSMC model, the partial energy deposition
rates for the accompanying flux of the primary photoelectrons caused by
the electron impact processes (5) in the H$_2 \rightarrow $H
transition region in the planetary upper atmosphere can be calculated. This
finally allows us to estimate the heating rate $W_T$ of the atmospheric gas by
photoelectrons in the planetary upper atmosphere and to calculate the heating-efficiency 
coefficient $\eta_{\rm h\nu}$, which is a critical parameter in the aeronomical
models~(\cite{Yelle-2008}). The heating efficiency $\eta_{\rm h\nu}$
is usually defined as a ratio of the absorbed energy accumulated as gas heat
to the deposited energy of the stellar radiation. We calculated the heating
efficiency in accordance with this definition, namely,
\begin{equation}
\eta_{\rm h\nu}(r) = \frac{W_{\rm T}(r)}{W_{\rm h\nu}(r)}.
 \end{equation}
In addition, the following simplified definition is sometimes used:
\begin{equation}
\eta_{\rm pe}(r) = \frac{W_{\rm pe}(r)}{W_{\rm h\nu}(r)},
 \end{equation}
which is an approximate ratio between kinetic energy stored by the
fresh (primary) photoelectrons and the deposited energy of the stellar radiation.

\subsection{Results}
We calculated the energy deposition of the stellar XUV radiation in
the H$_2 \rightarrow $H transition region ($1.04R_{\rm p}<R<1.2R_{\rm p}$) in the upper
atmosphere of HD 209458b. Height profiles of the main neutral constituents
H$_2$, H, and He were adopted from the aeronomical model of~\cite{Yelle-2004}.

We calculated the rate of the transition of the stellar XUV radiation and photoelectron energy
into the internal energy of the atmospheric gas in each of the photolytic and
electron-impact reactions. Additionally, the energy of
the suprathermal photoelectrons, which turns into heat was calculated.
Thus, the results of the simulation allow us to determine the total efficiency
of heating and heating efficiency by photoelectrons and to understand which
processes most affect the heating of the atmosphere.

\begin{figure}
\includegraphics[width=0.94\columnwidth]{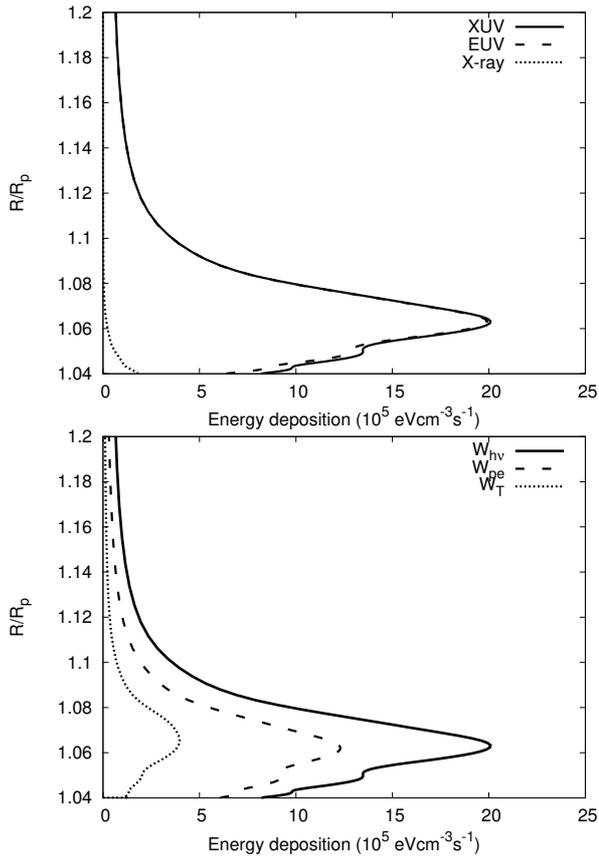}
\caption{
Top panel: Deposition rates due to the absorption of the stellar soft X-rays ($\lambda \sim$1--10 nm, dashed curve)
and the extreme ultraviolet radiation (EUV, $\lambda \sim$10--100 nm, dotted line) as well as
the total XUV radiation (solid line).
Bottom panel: Total deposition rate $W_{\rm h\nu}$ (solid curve) of the stellar XUV
radiation, the rate $W_{\rm pe}$ (dashed curve) of energy accumulation by the
fresh photoelectrons due to the photolytic processes~\eqref{1eq} and~\eqref{2eq},
and the heating rate $W_{\rm T}$ (dotted curve) by photoelectrons due to the
electron impact processes (5) in the H$_2 \rightarrow $H transition region
 in the upper atmosphere of HD 209458b.}
\label{2fig}
\end{figure}

\begin{figure}
\includegraphics[width=0.94\columnwidth]{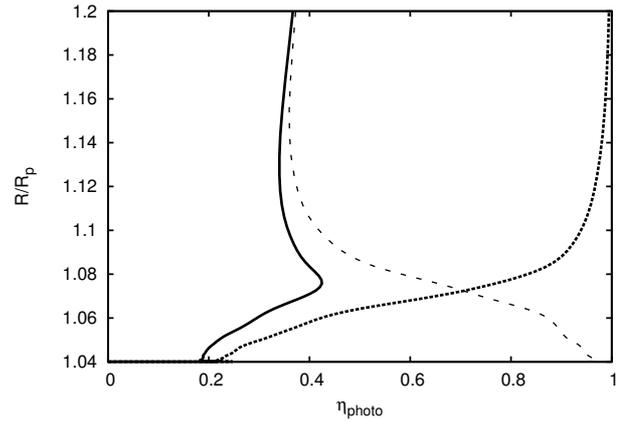}
\caption{Total $\eta_{\rm photo}(z) = \frac{W_{\rm T}(z)}{{W_{\rm pe}}(z)}$
and component-dependent heating efficiencies due to the electron impact processes (5)
in the H$_2\rightarrow$H transition region in the upper atmosphere of HD 209458b.
The solid line represents the total  $\eta_{photo}$, the dashed line shows the inner state excitation and ionization of H,
the dotted line represents the excitation and ionization of H$_2$ and He.}
\label{3fig}
\end{figure}

\begin{figure}
\includegraphics[width=0.94\columnwidth]{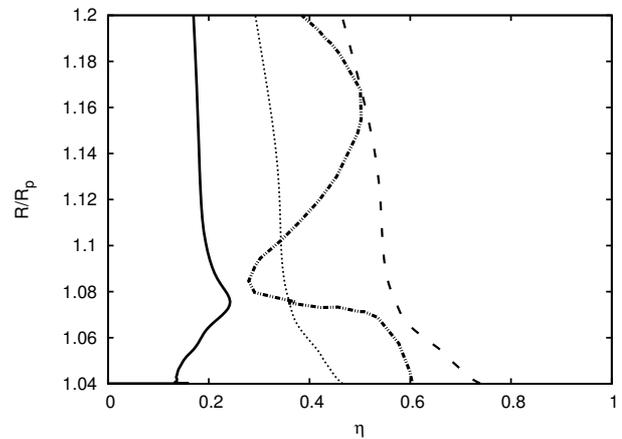}
\caption{Heating efficiency $\eta_{\rm h\nu}$ with (solid curve) and $\eta_{\rm pe}$
without (dashed curve) the photoelectron impact
processes~(5) in the H$_2 \rightarrow $H transition region in the upper
atmosphere of HD 209458b. For comparison the heating efficiency presented in Fig. 4 in the study by \cite{Yelle-2004}
is shown as the dotted-dashed line. The dotted line represents the heating efficiency $\eta_{\rm pe}$ reduced by a
factor 0.63 in accordance with the approach applied by \cite{Yelle-2004}.}
\label{4fig}
\end{figure}

In Fig.~\ref{2fig}~(upper panel) we show the deposition rates due to the absorption of the stellar soft X-rays ($\lambda \sim$1--10 nm) and the extreme ultraviolet radiation ($\lambda \sim$10--100 nm) as well as the total XUV radiation (solid line). The XUV-heating efficiency is dominated by the EUV photons. The input of X-ray photons becomes comparable with the EUV input at the very bottom boundary of the atmospheric region. In Fig.~\ref{2fig} (bottom panel) we show the total deposition rate $W_{\rm h\nu}$ (solid curve) of the stellar XUV
radiation, the rate $W_{\rm pe}$ (dashed curve) of energy accumulation by the
fresh photoelectrons due to the photo-ionization processes~\eqref{1eq} and~\eqref{2eq},
and the heating rate $W_{\rm T}$ (dotted curve) by photoelectrons due to the
electron impact processes (5) in the H$_2 \rightarrow $H transition region
in the upper atmosphere of HD 209458b.
The total $\eta_{\rm photo}(z)$ and component-dependent heating efficiencies due to the electron impact processes (5)
in the H$_2\rightarrow$H transition region in the upper atmosphere of HD 209458b are
given in Fig.~\ref{3fig}.  The solid curve shows the height profile of the ratio
$\eta_{\rm photo}(z) = \frac{W_{\rm T}(z)}{{W_{\rm pe}}(z)}$, which could
be considered as the heating efficiency due to the electron impact processes (5) alone.
Both excitation of internal states and the ionization of molecular
and atomic hydrogen are the dominant channels of the photoelectron energy
deposition.
The values shown in Figs.~\ref{2fig} and ~\ref{3fig} allow us to calculate the heating
efficiency $\eta_{\rm h\nu}$ defined as the ratio of the absorbed energy accumulated as gas heating
to the deposited energy of the stellar XUV radiation, namely
\begin{equation}
\eta_{\rm h\nu}(r)=\eta_{\rm pe}(r) \times \eta_{\rm photo}(r).
 \end{equation}
Fig.~\ref{4fig} shows the heating efficiency $\eta_{\rm h\nu}$ with (solid curve) and $\eta_{\rm pe}$ without (dashed curve)
the photoelectron impact processes (5) in the H$_2 \rightarrow $H transition region
in the upper atmosphere of HD 209458b. The total heating efficiency is height-dependent with
the values varying in the range $\sim$10\%--25\% and with a peak value between $\sim$1.07--1.08$R_{\rm p}$ that
approaches 25\%.
A comparison of our modeled heating efficiency with that of \cite{Yelle-2004}
is shown as the dotted-dashed line in Fig.~\ref{4fig}. In the aeronomic model of \citep{Yelle-2004}) the heating rate caused by the fresh
photoelectrons was not calculated in detail. It was assumed that the extra energy acquired by the photoelectron
in an ionization event is transferred to the ambient atmosphere with an efficiency of $\sim$63\%, that is, the heating efficiency
$\eta_{\rm h\nu}$ was approximated as equal to 0.63$\times\eta_{\rm pe}$. Therefore, we also show in Fig.~\ref{4fig} as a dotted line
our calculated heating efficiency $\eta_{\rm pe}$ (dashed line) reduced by factor 0.63 in accordance with the approach used by Yelle (2004).
Clearly, this approach still overestimates the heating rate of the ambient atmospheric gas by the photoelectrons.

Our results indicate that the effect of the electron-impact processes shown in Eq. (5) together with the participation of
the suprathermal photoelectrons play an important role in heating the upper atmosphere of hydrogen-dominated exoplanets
by the stellar XUV radiation and should not be neglected. The height profiles of the heating efficiency by the stellar XUV radiation were calculated, and its value does not exceed 20\% almost everywhere in the $H_2 \rightarrow H$ transition region of the hydrogen-rich thermosphere. Thus, the correct account of photoelectrons reduces $\eta_{\rm h\nu}$ by $\sim$3--4 times.

Our model $\eta_{\rm h\nu}$ values for
hydrogen-dominated exoplanet upper atmospheres are similar to the atmosphere of Jupiter
in the solar system. Waite et al. (1983) have shown that the average heating efficiency $\eta_{\rm h\nu}$ value of the solar EUV radiation for
neutral gas in the hydrogen dominated thermosphere of Jupiter is $\sim$9.26 \%.

Our results are quite relevant especially for statistical evolutionary
atmospheric mass-loss studies of transiting exoplanets. Based on the idea
of Lecavelier des Etangs (2007), who produced an energy diagram by comparing
the stellar XUV energy received by the upper atmospheres to the gravitational
energies of exoplanets, but by introducing the heating efficiency $\eta_{\rm h\nu}$ as in Lammer et al. (2009) within a range between 1--100 \%,
Ehrenreich and D\'{e}sert (2011) estimated atmospheric mass-loss rates
during the lifetimes of close-in exoplanets. 

These authors proposed to estimate $\eta_{\rm h\nu}$ when the mass-loss power $L$ can be
constrained by exoplanet transit observations in Ly-$\alpha$ (e.g., Vidal-Madjar et al. 2003; Lecavelier des Etangs et al. 2010)
and the stellar XUV luminosity $L_{\rm XUV}$ is known. By using the relation $\eta_{\rm h\nu}=L/L_{\rm XUV}$ Ehrenreich and 
D\'{e}sert (2012) obtained physically impossible $\eta_{\rm h\nu}$ values of $>$100 \% for HD209458b and $\sim$1 \% for
HD189733b. To explain the unphysical high $\eta_{\rm h\nu}$ value obtained for  HD209458b other hypotheses such as
the sudden sporadic change in $L_{\rm XUV}$ have been proposed.

However, as shown in our detailed study, the XUV-related $\eta_{\rm h\nu}$ value should be within 10--20\% and will therefore not reach the high values necessary to reproduce the observation-based mass-loss rates for HD 209458b with the
hypothesis presented in Ehrenreich and D\'{e}sert (2011). The most likely reason why the hypothesis of Ehrenreich and D\'{e}sert (2011) will not give accurate $\eta_{\rm h\nu}$ values is related to the fact that similarly to Lecavelier des Etangs (2007) or Lammer et al. (2009), they assumed that the effective XUV absorption radius $R_{\rm XUV}$ lies close to $R_{\rm p}$, the radius used in the energy-limited mass-loss formula for all of the studied exoplanets. This assumption is more or less valid for massive and compact exoplanets (Erkaev et al. 2007), such as HD 189733b with an average density $\rho \sim 0.95$ g cm$^{-3}$, but will yield less accurate mass-loss rates for less compact objects with lower average densities, such as HD 209458b with 0.37 g cm$^{-3}$. This effect was first recognized by Watson et al. (1981) and was more recently investigated in detail by Erkaev et al. (2013; 2014). $R_{\rm XUV}$ can exceed the planetary radius $R_{\rm p}$ quite substantially for a planetary body with a low average density when its atmosphere is exposed to high XUV fluxes. Depending on the distribution of the XUV volume-heating rate and the related density profile of the upper atmosphere, the mass-loss rate can then be higher, as estimated with the assumptions in Ehrenreich and D\'{e}sert (2011).

Because an $R_{\rm XUV}$ that is larger than $R_{\rm p}$ can only be estimated by applying an XUV absorption and hydrodynamic upper atmosphere model, mass-loss studies that apply the energy-limited formula and assume $R_{\rm XUV}\approx R_{\rm p}$ may underestimate the loss rates, even after the modification by an accurate heating efficiency $\eta_{\rm h\nu}$ of $\leq$20\%.

\section{Conclusion}
We modeled the heating efficiency of the stellar XUV radiation within a hydrogen-dominated upper atmosphere, such as that of
the extra-solar gas giant HD 209458b. We showed that one cannot neglect the effects caused by
electron-impact processes together with the participation of suprathermal photoelectrons. By including these
processes, our Direct Simulation Monte Carlo model results indicate that the XUV heating efficiency $\eta_{\rm h\nu}$ in
hydrogen envelopes or hydrogen-dominated planetary upper atmospheres approaches 10 \% at the bottom boundary $R$ =1.04$R_p$ of the region and its value does not exceed the value 20 \% over most of the thermosphere. Our result agrees well
with the heating-efficiency values assumed in early studies by \cite{Kasting-and-Pollack-1983} and \cite{Chassefiere-1996}. Therefore, we conclude that atmospheric mass-loss studies that assumed $\eta_{\rm h\nu}$ values $>$ 20\% overestimated the hydrogen-escape rates, while hydrogen-escape studies that assumed $\eta_{\rm h\nu}$ values $\sim$10--15\% probably yield accurate results.

\begin{acknowledgements}
H. Lammer acknowledges the support by the FWF NFN project S116 ``Pathways to Habitability: From Disks to Active Stars, Planets and Life'', and the related FWF NFN subproject, S116 607-N16 ``Particle/Radiative Interactions with Upper Atmospheres of Planetary Bodies Under Extreme Stellar Conditions''. V.I. Shematovich and D.E. Ionov acknowledge the support by the Russian Science Foundation (Project no. 14-12-01048). The authors also thank the International Space Science Institute (ISSI) in Bern and the ISSI team ``Characterizing stellar and exoplanetary environments.''
\end{acknowledgements}

\end{document}